\documentclass[12pt]{article}
\usepackage{veniceproc}

\usepackage{graphicx,epsfig,a4,fancybox,fancyhdr,epic,rotating}

\begin{document}

\begin{center}

{\large\bf NEUTRINO OSCILLATIONS}
\\
\vspace*{0.1cm}

{\bf A Historical Overview and its Projection} \\
\vspace{0.5cm}

PETER MINKOWSKI\footnote{\hspace*{0.1cm} Work supported in part by
Schweizerischer Nationalfonds.}  \\[3mm] {\it ITP, University of Bern,
Switzerland} \\
\vspace{0.5cm}
\end{center}

\begin{center}
{\large \bf Topics}
\end{center}
\vspace*{0.3cm}

\hspace*{1.0cm} 1. Base fermions and scalars in SO10 
\vspace*{+0.1cm}

\hspace*{1.4cm}
neutrinos are unlike charged fermions - Ettore Majorana
\vspace*{0.1cm}

\hspace*{1.0cm} 2. Neutrino  
'mass from mixing' in vacuo and matter 
\vspace*{+0.1cm}

\hspace*{1.4cm}
neutrinos oscillate like neutral Kaons 
\vspace*{+0.1cm}

\hspace*{1.4cm}
(yes, but how ?)  - Bruno Pontecorvo
\vspace*{+0.1cm}

\hspace*{1.0cm} 3.  Some perspectives

\section{Charged fermions are not like neutrinos \footnote{\hspace*{0.1cm}
[1] Ettore Majorana, ``Teoria simmetrica dell'elettrone'', Nuovo Cimento 14
(1937) 171.}}

We shall consider -'pour fixer les idees'- 3 fermion families in the (left-)
chiral basis, forming a substrate for the local gauge group 
SL(2,C) [or SO(1,3)] x SO10

\begin{center}
\begin{figure}[htb]
\epsfig{file=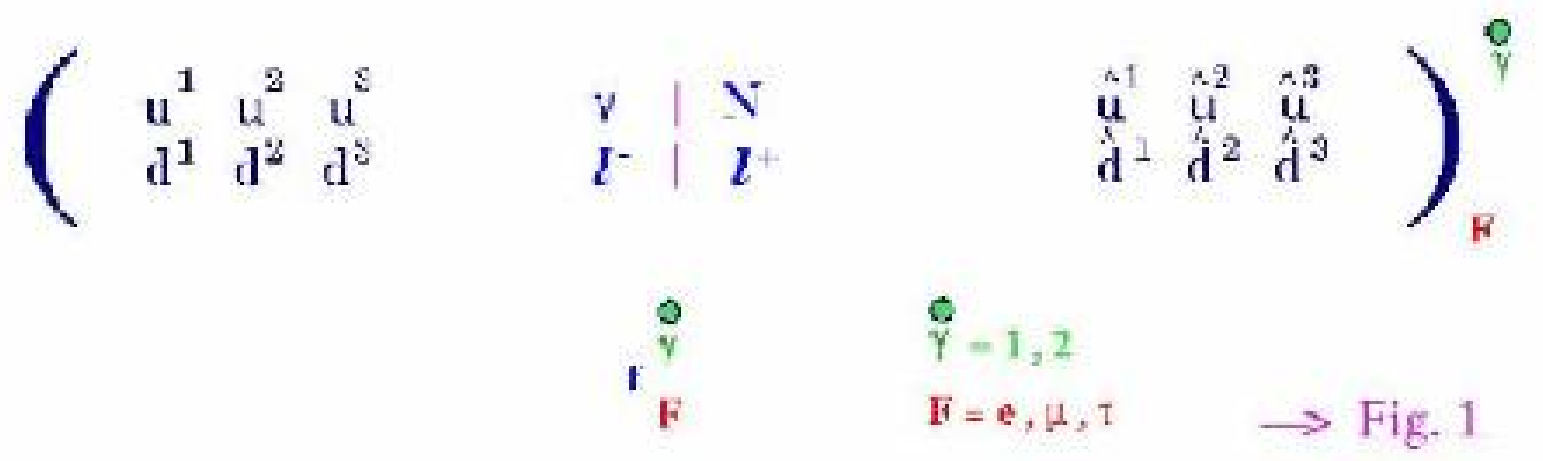,width=13cm}
\vskip -0.2cm
 \end{figure}
\end{center}

\noindent \underline{Key questions $\rightarrow$ why 3? Why SO10?}  I shall
cite two sentences from ref.\ [1]: ``Per quanto riguarda gli elettroni e i
positroni, da essa (via) si pu\`{o} veramente attendere soltanto un
progresso formale...  Vedremo infatti che \`{e} perfettamente possibile
costruire, nella maniera pi\`{u} naturale, una teoria delle particelle
neutre elementari senza stati negativi.''  (``As far as electrons and
positrons are concerned from this (path) one may expect only a formal
progress...  We will see in fact that it is perfectly possible to construct,
in the most general manner, a theory of neutral elementary particles without
negative states.'') \footnote{\hspace*{0.1cm} ... upon normal ordering.}

\vspace*{0.3cm}

\noindent
But the real content of the paper by E.\ M.\ (1937) is in the formulae,
exhibiting the 'oscillator decomposition' of spin 1/2 fermions as seen and
counted by gravity, 1 by 1 and doubled through the 'external' SO2 symmetry
associated with electric charge [ 2 ].
\footnote{\hspace*{0.1cm} [ 2 ] P.\ A.\ M.\ Dirac, Proceedings of the
Cambridge Philosophical Society 30 (1924) 150. Paul Dirac shall be excused
for starting the count at 2 for 'elettrone e positrone' .}
\vspace*{0.2cm}

\noindent
The left chiral notation shall be
\vspace*{0.4cm}

\begin{equation}
\label{nu's:1}
\begin{array}{l}
\left ( \ f_{\ k} \ \right )_{\ F}^{\ \dot{\gamma}}
\hspace*{0.1cm} ; \hspace*{0.1cm} ^{\dot{\gamma}} \ = \ 1,2 
\hspace*{0.1cm} : \hspace*{0.1cm} \mbox{spin projection} 
\vspace*{0.3cm} \\ 
\hspace*{1.9cm} _{\ F} \ = \ I,II,III
\hspace*{0.1cm} : \hspace*{0.1cm} \mbox{family label}
\vspace*{0.3cm} \\ 
\hspace*{1.9cm} _{\ k} \ = \ 1, \cdots , 16
\hspace*{0.1cm} : \hspace*{0.1cm} \mbox{SO10 label}
\end{array}
\end{equation}
\vspace*{0.0cm}

\noindent
Lets call the above extension of the standard model the 'minimal nu-extended
SM' [ 3 ]. \footnote{
\hspace*{0.2cm} [ 3 ] Harald Fritzsch and Peter Minkowski, ``Unified
interactions of leptons and hadrons'', Annals Phys.\ 93 (1975) 193 and
Howard Georgi, ``The state of the art -- gauge theories'', AIP Conf.\ Proc.\
23 (1975) 575.}

\vspace*{-0.5cm}
\begin{equation}
\label{nu's:2}
\begin{array}{c}
\left (
\ \begin{array}{lll lll lll}
\bullet & \bullet & \bullet & \nu & | 
& {\cal{N}} & \bullet & \bullet & \bullet
\vspace*{0.3cm} \\
\bullet & \bullet & \bullet & \ell & | 
& \widehat{\ell} & \bullet & \bullet & \bullet
\ \end{array}
\ \right )_{\ F \ = \ e \ , \ \mu \ , \ \tau}^{\ \dot{\gamma}} 
\vspace*{0.3cm} \\
\hspace*{-2.6cm} \downarrow
\vspace*{0.2cm} \\
\hspace*{-0.4cm} \left (
\ \begin{array}{lll}
\nu & & {\cal{N}} 
\vspace*{0.3cm} \\
\ell & & \widehat{\ell} 
\ \end{array}
\ \right )_{\ F \ = \ e \ , \ \mu \ , \ \tau}^{\ \dot{\gamma}}
\end{array}
\end{equation}
\vspace*{0.2cm}

\noindent
The right-chiral base fields are then associated to
\hspace*{0.1cm} ( 1 for 1 )

\vspace*{-0.0cm}
\begin{equation}
\label{nu's:3}
\begin{array}{l}
\left ( \ f^{\ *}_{\ k} \ \right )_{\ F \ \alpha}
\ = \ \varepsilon_{\ \alpha \gamma}
\ \left \lbrack \ \left ( \ f_{\ k} \ \right )_{\ F}^{\ \dot{\gamma}}
\ \right \rbrack^{\ *}
\vspace*{0.3cm} \\
\left ( \ \varepsilon \ = \ i \ \sigma_{\ 2} \ \right )_{\ \alpha \gamma}
\ = \ \left (
 \begin{array}{rl}
0 & 1
\vspace*{0.3cm} \\
- 1 & 0
\end{array}
\ \right ) 
\end{array}
\end{equation}
\vspace*{0.2cm}

\noindent
The matrix $\varepsilon$ is the symplectic ($Sp\ (1)$) unit, as implicit in
Ettore Majorana's original paper [ 1 ].

\noindent
The local gauge theory is based on the gauge (sub-) 
group

\vspace*{-0.1cm}
\begin{equation}
\label{nu's:4}
\begin{array}{l}
SL \ ( \ 2 \ , \ C \ ) \ \times \ SU3_{\ c} \ \times
\ SU2_{\ L} \ \times \ U1_{\ {\cal{Y}}}
\end{array}
\end{equation}

\noindent
... why ? why 'tilt to the left' ?
we sidestep 
a historical overview here !
\vspace*{0.3cm}

\section{Yukawa interactions and mass terms} 

The doublet(s) of scalars are related to the 'tilt to the left' . 

\vspace*{-0.0cm}
\begin{equation}
\label{nu's:5}
\begin{array}{l}
\left (
\ \begin{array}{lll}
\nu & & {\cal{N}} 
\vspace*{0.3cm} \\
\ell & & \widehat{\ell} 
\ \end{array}
\  \right )_{\vspace*{1.4cm} \hspace*{-0.2cm}
\begin{array}{l} {\scriptstyle F}
\end{array}}
\ \leftrightarrow
\ \left (
\ \begin{array}{lll}
\varphi^{\ 0} & & \Phi^{\ +} 
\vspace*{0.3cm} \\
\varphi^{\ -} & & \Phi^{\ 0}  
\ \end{array}
 \right )
\ = \ z
\end{array}
\end{equation}
\vspace*{0.2cm}

\noindent
The entries ${\cal{N}} \ , \ \widehat{\ell}$ in eq. (5) denote singlets 
under $SU2_{\ L}$ .
\vspace*{0.1cm}

\noindent
The quantity z is associated with the quaternionic or octonionic structure
inherent to the $( \ 2 \ , \ 2 \ )$ representation of $SU2_{\ L} \ \otimes \
SU2_{\ R}$ (beyond the electroweak gauge group)\footnote{
\hspace*{0.2cm} e.g.\ [ 4 ] F.\ G\"{u}rsey and C.\ H.\ Tze, ``On the role of
division- , Jordan- and related algebras in particle physics'', Singapore,
World Scientific (1996) 461.}.
\vspace*{0.2cm}

\noindent
The Yukawa couplings are of the form
(notwithstanding the quaternionic or octonionic structure of
scalar doublets)

\vspace*{-0.0cm}
\begin{equation}
\label{nu's:6}
\begin{array}{l}
{\cal{H}}_{\ Y} \ =
\ \left \lbrack 
\ ( \ \varphi^{\ 0} \ )^{\ *} \ , \ ( \ \varphi^{\ -} \ )^{\ *}
\ \right \rbrack
\ \lambda_{ \ F' \ F} \ \times
\vspace*{0.3cm} \\
\hspace*{2.2cm} 
\ \times 
\ \left \lbrace 
\ \varepsilon_{\dot{\gamma} \dot{\delta}}
\ {\cal{N}}^{\ \dot{\delta}}_{\ F'}
\ \left \lbrack
\begin{array}{l}
 \nu^{\ \dot{\gamma}}
\vspace*{0.3cm} \\
\ell^{\ \dot{\gamma}}
\end{array}
\ \right \rbrack_{\ F}
\ \right \rbrace \ + \ h.c.
\vspace*{0.5cm} \\
{\cal{N}}_{\ \dot{\gamma} \ F'} \ =
\ \varepsilon_{\dot{\gamma} \dot{\delta}}
\ {\cal{N}}^{\ \dot{\delta}}_{\ F'}
\hspace*{0.2cm} ; \hspace*{0.2cm}
\varepsilon_{\dot{\gamma} \dot{\delta}} \ = 
\ \overline{\varepsilon_{\gamma \delta}} 
\ = \ \varepsilon_{\gamma \delta} 
\end{array}
\end{equation}
\vspace*{0.2cm}

\noindent
The only allowed Yukawa couplings by $SU2_{\ L} \ \otimes \ U1_{\
{\cal{Y}}}$ invariance are those in eq. (6) , with arbitrary complex
couplings $\lambda_{\ F' \ F}$.  Spontaneous breaking of $SU2_{\ L} \
\otimes \ U1_{\ {\cal{Y}}}$ through the vacuum expected value(s)

\vspace*{-0.0cm}
\begin{equation}
\label{nu's:8}
\begin{array}{l}
\left \langle \ \Omega \ \right |
\ \left (
\ \begin{array}{lll}
\varphi^{\ 0} & & \Phi^{\ +} 
\vspace*{0.3cm} \\
\varphi^{\ -} & & \Phi^{\ 0} 
\ \end{array}
 \right ) \ ( \ x \ )
\ \left | \ \Omega \ \right \rangle
\ = 
\vspace*{0.3cm} \\
 =  \left \langle \ z \ ( \ x \ ) \ \right \rangle
\ = \ \left (
\ \begin{array}{lll}
v_{\ ch} \ (v^{\ u}_{\ ch}) & &  0 
\vspace*{0.3cm} \\
0 & & v_{\ ch}  \  (v^{\ d}_{\ ch})
\ \end{array}
 \right )
\vspace*{0.3cm} \\
v_{\ ch} \ = \ \frac{1}{\sqrt{2}}
\ \left ( \ \sqrt{2} \ G_{\ F} \ \right )^{\ -1/2}
\ = \ 174.1 \ \mbox{GeV}
\end{array}
\end{equation}

\noindent
independent of the space-time point $x$ 
\footnote{\hspace*{0.2cm} The implied parallelizable
nature of $\left \langle \ z \ ( \ x \ ) \ \right \rangle$
is by far not trivial and relates in a wider context including
triplet scalar representations to potential
(nonabelian) monopoles and dyons. (no h.o.)} ,
induces a neutrino mass
term through the
Yukawa couplings $\lambda_{\ F' \ F}$ in
eq. (6)

\vspace*{-0.0cm}
\begin{equation} 
\label{nu's:8a}
\begin{array}{c}
_{\ F'} \ {\cal{N}} \ \nu_{\ F} \ =
\ {\cal{N}}_{\ \dot{\gamma} \ F'} \ \nu^{\ \dot{\gamma}}_{\ F}
\ = \ \nu_{\ \dot{\gamma} \ F} \ {\cal{N}}^{\ \dot{\gamma}}_{\ F'}
\vspace*{0.2cm} \\ 
\mu_{\ F' \ F} \ = \ v_{\ ch} \ \lambda_{\ F' \ F}
\vspace*{0.2cm} \\ 
\rightarrow \ {\cal{H}}_{\ \mu} \ =
\ _{\ F'} \ {\cal{N}} \ \mu_{\ F' \ F} \ \nu_{\ F} \ + \ h.c.
\ = \ \nu^{\ T} \ \mu^{\ T} \ {\cal{N}} \ + \ h.c.
\end{array}
\end{equation}

\noindent
The matrix $\mu$ defined in eq. (8) is an arbitrary
complex $3 \ \times \ 3$ matrix, analogous to the
similarly induced mass matrices of charged 
leptons and quarks.
In the setting of primary SO10 breakdown, a general (not symmetric)
Yukawa coupling $\lambda_{\ F' \ F}$ implies the existence
in the scalar sector of at least two irreucible representations
$(16) \oplus (120)$ \footnote{\hspace*{0.2cm} 
key question $\rightarrow$ 
a 'drift' towards unnatural complexity  ?
It becomes even worse including the heavy neutrino mass terms :
256 (complex) scalars.}.

\section{Mass from mixing' in vacuo [ 5 -- 7 ], or 'Seesaw' [ 8 -- 11 ]; 
neutrinos oscillate like neutral Kaons (yes, but how?)  -- Bruno
Pontecorvo\footnote{\hspace*{0.2cm} We will come back to the clearly
original idea {\bf in 1957} of Bruno Pontecorvo [ 12 - ] but let me first
complete the 'flow of thought' embedding neutrino masses in SO10.}}

\noindent
The special feature, pertinent to (electrically neutral)
neutrinos is, that the $\nu-$ extending degrees of freedom
${\cal{N}}$ are
singlets under the whole SM gauge group
$G_{\ SM} \ = 
\ SU3_{\ c} \ \otimes \ SU2_{\ L} \ \otimes \ U1_{\ {\cal{Y}}}$ ,
in fact remain singlets under
the larger gauge group $SU5 \ \supset \ G_{\ SM}$ .
This allows an arbitrary (Majorana-) mass term, involving 
the bilinears formed from two ${\cal{N}}$-s.

\noindent
In the present setup (minimal $\nu$-extended SM) the full neutrino
mass term is thus of the form

\vspace*{-0.0cm}
\begin{equation}
\label{nu's:9}
\begin{array}{l}
{\cal{H}}_{\ {\cal{M}}} \ =
\ \frac{1}{2} \ \left \lbrack \ \nu \ {\cal{N}} \ \right \rbrack
\ {\cal{M}} \ \left \lbrack
\begin{array}{l}
\nu
\vspace*{0.3cm} \\
{\cal{N}}
\end{array}
\ \right \rbrack
\ + \ h.c.
\vspace*{0.3cm} \\
{\cal{M}} \ =
\ \left (
 \begin{array}{ll}
 0 &  \mu^{\ T}
\vspace*{0.3cm} \\
 \mu &  \ M 
\end{array}
 \right )
\hspace*{0.2cm} ; \hspace*{0.2cm}
 {\cal{M}} \ = \ {\cal{M}}^{\ T} \ \rightarrow
\hspace*{0.1cm} 
 M \ = \ M^{\ T} 
\end{array}
\end{equation}
\vspace*{0.3cm}

\noindent
Again within primary SO10 breakdown
the full ${\cal{M}}$
extends the scalar sector to the representations
$(16) \ \oplus \ (120) \ \oplus \ (126)$ 
\footnote{\hspace*{0.2cm} It is from here 
where the discussion
-- to the best of my knowledge --
of the origin and magnitude of the light neutrino masses (re-) started
in 1974 as documented here.} 
. \vspace*{0.3cm}

\noindent
[ 5 ] Harald Fritzsch, Murray Gell-Mann and Peter Minkowski,
``Vector-like weak currents and new elementary fermions'',
Phys.\ Lett.\ B59 (1975) 256.
\vspace*{0.1cm}

\noindent
[ 6 ] Harald Fritzsch and Peter Minkowski, ``Vector-like weak currents,
massive neutrinos, and neutrino beam oscillations'', Phys.\ Lett.\ B62
(1976) 72.  ([ 5 ] and [ 6 ] in the the general vector-like situation.)
\vspace*{0.1cm}

\noindent
and for 'our world, tilted to the left'  
\vspace*{0.2cm}

\noindent
$\left \lbrack \ 7 \ \right \rbrack$
Peter Minkowski,
``$\mu \ \rightarrow \ e \gamma$ at a rate of one out of
1-billion muon decays?'',
Phys.\ Lett.\ B67 (1977) 421. 

\vspace*{0.2cm}

\noindent 
Correct derivations were subsequently {\it documented}
in $\left \lbrack \ 8 \ \right \rbrack$ -
$\left \lbrack \ 11 \ \right \rbrack$:
\vspace*{0.1cm}

\noindent
$\left \lbrack \ 8 \ \right \rbrack$ Murray Gell-Mann, Pierre Ramond and
Richard Slansky, ``Complex spinors and unified theories'', published in
Supergravity, P.\ van Nieuwenhuizen and D.Z.\ Freedman (eds.), North Holland
Publ.\ Co., 1979 and in Stony Brook Wkshp.\ 1979:0315 (QC178:S8:1979).

\noindent
$\left \lbrack \ 9 \ \right \rbrack$ Tsutomu Yanagida, ``Horizontal symmetry
and masses of neutrinos'', published in the Proceedings of the Workshop on
the Baryon Number of the Universe and Unified Theories, O.\ Sawada and A.\
Sugamoto (eds.), Tsukuba, Japan, 13-14 Feb.\ 1979, and in (QCD161:W69:1979)
.

\noindent
$\left \lbrack \ 10 \ \right \rbrack$
Shelley Glashow,
``Quarks and leptons'', published in Proceedings
of the Carg\`{e}se Lectures, M.\ L\'{e}vy (ed.), Plenum Press,
New York, 1980.  

\noindent
$\left \lbrack \ 11 \ \right \rbrack$
Rabindra Mohapatra and Goran Senjanovi\v{c},
``Neutrino mass and spontaneous parity violation'',
Phys.\ Rev.\ Lett.\ 44 (1980) 912.

\noindent
We resume the discussion of the mass term in eq. (9).

\noindent
Especially the $0$  entry needs explanation.
It is an exclusive property of the minimal $\nu$-extension assumed here.
\vspace*{0.2cm}

\noindent
Since the 'active' flavors $\nu_{\ F}$ all carry
$I_{\ 3 \ w} \ = \ \frac{1}{2}$ terms of the form

\vspace*{-0.0cm} 
\begin{equation}
\label{nu's:10}
\begin{array}{l}
\frac{1}{2} \  _{\ F'} \ \nu \ \chi_{\ F' \ F} \ \nu_{\ F}
\ = \ \frac{1}{2} \ \nu^{\ T} \ \chi \ \nu
\hspace*{0.2cm} ; \hspace*{0.2cm}
\chi \ = \ \chi^{\ T}
\end{array}
\end{equation}

\noindent
cannot arise as Lagrangean masses, except induced by an
$I_{\ w}$-triplet of scalars, developing a vacuum expected
value independent from the doublet(s)
\footnote{\hspace*{0.1cm} key questions $\rightarrow$ quo vadis ?  is this a
valid explanation of the 'tilt to the left' ? no, at least insufficient!}.

\noindent
from ``The apprentice magician'' by Goethe :
\vspace*{0.1cm}

\noindent
'The shadows I invoked, I am unable to get rid of now !'

\section{Neutrino oscillations - historical overview}

\noindent
The idea that light neutrinos have mass and oscillate goes back to
Bruno Pontecorvo, but starting
with (para-) muonium - antimuonium 
oscillations $\left \lbrack \ 12 \ \right \rbrack$ 
\footnote{\hspace*{0.1cm} $\left \lbrack \ 12 \ \right \rbrack$ Bruno
Pontecorvo, ``Mesonium and antimesonium'', JETP (USSR) 33 (1957) 549,
english translation Soviet Physics, JETP 6 (1958) 429.} - like $K^{\ 0} \
\leftrightarrow \ \overline{K}_{\ 0}$ $\left \lbrack \ 13 \ \right \rbrack$
\footnote{\hspace*{0.1cm} 
$\left \lbrack \ 13 \ \right \rbrack$
Murray Gell-Mann and Abraham Pais, Phys.\ Rev.\ 96 (1955) 1387,
introducing $\tau$.}.

\noindent
Assuming CP conservation there are two equal mixtures of $\mu^{\ -} \ e^{+}$
and $\mu^{\ +} \ e^{-}$ with opposite CP values $\pm$
(at rest and using a semiclassical description of quantum states)

\vspace*{-0.1cm}
\begin{equation}
\label{nu's:11}
\hspace*{-0.3cm}
\begin{array}{l}
\left | \ ( \ e \mu \ )_{\ \pm} \ ; \ \tau  =  0 \ \right \rangle
\ = \ \frac{1}{\sqrt{2}}
\ \left (
\ \left | \ ( \ e^{\ -} \mu^{\ +} \ ) \ \right \rangle
\ \mp
\ \left | \ ( \ e^{\ +} \mu^{\ -} \ ) \ \right \rangle
\ \right )_{\ \tau  =  0}
\vspace*{0.3cm} \\
\left | \ ( \ e^{\ +} \mu^{\ -} \ ) \ \right \rangle
\ = \ \widehat{C} 
\ \left | \ ( \ e^{\ -} \mu^{\ +} \ ) \ \right \rangle
\end{array}
\end{equation}

\noindent
For the leptonium case the rest system is a good appoximation.

\noindent
The evolution of the CP $\pm$ states is then characterized by

\vspace*{-0.1cm}
\begin{equation}
\label{nu's:12}
\hspace*{0.3cm}
\begin{array}{l}
\widehat{m}_{\ \alpha} \ = \ m_{\ \alpha} \ - \ \frac{i}{2} \ \Gamma_{\ \alpha} 
\hspace*{0.2cm} ; \hspace*{0.2cm} 
\alpha \ = \ \pm
\hspace*{0.2cm} \mbox{with} \hspace*{0.2cm}
\vspace*{0.3cm} \\
\left | \ ( \ e^{\ -} \mu^{\ +} \ ) \ \right \rangle \ =
\ \left | \ 1 \ \right \rangle
\ \rightarrow_{\ \tau}
\vspace*{0.3cm} \\
\ \frac{1}{\sqrt{2}} 
\ \left (
\ \left | \ + \ ; \ \tau  =  0 \ \right \rangle 
\ e^{\ - i \ \widehat{m}_{\ +} \ \tau}
\ +
\ \left | \ - \ ; \ \tau  =  0 \ \right \rangle 
\ e^{\ - i \ \widehat{m}_{\ -} \ \tau}
\ \right )
\vspace*{0.3cm} \\
\left | \ ( \ e^{\ +} \mu^{\ -} \ ) \ \right \rangle \ =
\ \left | \ 2 \ \right \rangle 
\ \rightarrow_{\ \tau}
\vspace*{0.3cm} \\
\ \frac{1}{\sqrt{2}} 
\ \left ( 
\ - \ \left | \ + \ ; \ \tau  =  0 \ \right \rangle 
\ e^{\ - i \ \widehat{m}_{\ +} \ \tau}
\ +
\ \left | \ - \ ; \ \tau  =  0 \ \right \rangle 
\ e^{\ - i \ \widehat{m}_{\ -} \ \tau}
\ \right ) 
\end{array}
\end{equation}
\vspace*{0.2cm}

\noindent
This reconstructs to 

\vspace*{-0.1cm}
\begin{equation}
\label{nu's:13}
\hspace*{0.3cm}
\begin{array}{l}
\left | \ 1 \ \right \rangle
\ \rightarrow_{\ \tau}
\hspace*{0.8cm} E_{\ +} \ ( \ \tau \ ) \ \left | \ 1 \ \right \rangle \ -
\ E_{\ -} \ ( \ \tau \ ) \ \left | \ 2 \ \right \rangle
\vspace*{0.3cm} \\
\left | \ 2 \ \right \rangle
\ \rightarrow_{\ \tau}
\hspace*{0.4cm} - \ E_{\ -} \ ( \ \tau \ ) \ \left | \ 1 \ \right \rangle \ +
\ E_{\ +} \ ( \ \tau \ ) \ \left | \ 2 \ \right \rangle
\vspace*{0.3cm} \\
 E_{\ \pm} \ ( \ \tau \ ) \ = \ \frac{1}{2}
\ \left (
\ e^{\ - i \ \widehat{m}_{\ +} \ \tau} \ \pm
\ e^{\ - i \ \widehat{m}_{\ -} \ \tau}
\ \right )
\end{array}
\end{equation}
\vspace*{0.2cm}

\noindent
and leads to the transition {\it relative}
probabilities indeed identical to the

\noindent
$K^{\ 0} \ \rightarrow \ \left | \ 1 \ \right \rangle \ ;
\overline{K}_{\ 0} \ \rightarrow \ \left | \ 2 \ \right \rangle$
system.

\vspace*{-0.1cm}
\begin{equation}
\label{nu's:14}
\hspace*{0.3cm}
\begin{array}{l}
d \ p_{\ 1 \ \leftarrow \ 1} \ = \ d \ p_{\ 2 \ \leftarrow \ 2}
\ = \ \left | \ E_{\ +} \ ( \ \tau \ ) \ \right |^{\ 2} \ d \ \tau
\vspace*{0.3cm} \\
d \ p_{\ 2 \ \leftarrow \ 1} \ = \ d \ p_{\ 1 \ \leftarrow \ 2}
\ = \ \left | \ E_{\ -} \ ( \ \tau \ ) \ \right |^{\ 2} \ d \ \tau
\vspace*{0.3cm} \\
\left | \ E_{\ \pm} \ ( \ \tau \ ) \ \right |^{\ 2} \ =
\vspace*{0.2cm} \\
\ = \ \frac{1}{2} 
\ e^{\ - \frac{1}{2} \ ( \ \Gamma_{\ +} \ + \ \Gamma_{\ -} \ ) \ \tau}
\ \left ( \ \cosh \ \frac{1}{2} \ \Delta \ \Gamma \ \tau 
\ \pm \ \cos \ \Delta \ m \ \tau
\ \right )
\hspace*{0.3cm} \rightarrow
\end{array}
\end{equation}

\noindent
The term $\cos \ \Delta \ m \ \tau$ 
in eq. (14) indeed signals 
$( \ e \widehat{\mu} \ ) \ \leftrightarrow \ ( \ \widehat{e} \mu \ )$
oscillations, with 

\vspace*{-0.1cm}
\begin{equation}
\label{nu's:15}
\hspace*{0.4cm}
\begin{array}{l}
\Delta \ m \ = 
\ m_{\ +} \ - \ m_{\ -}
\hspace*{0.2cm} ; \hspace*{0.2cm}
\Delta \ \Gamma \ = \ \Gamma_{\ +} \ - \ \Gamma_{\ -}
\vspace*{0.3cm} \\ 
 \Delta \ m \ = \ O
\ \left \lbrack \ \left (
\ \begin{array}{c}
\alpha \ m_{\ e} \ m_{\ \nu_{e} \nu_{ \mu}}
\vspace*{0.2cm} \\ \hline \vspace*{-0.4cm} \\
v^{\ 2}
\end{array}
\ \right )^{\ 2} \ m_{\ \mu} 
\ \right \rbrack
\ \sim \ 4 \ . \ 10^{\ -41} \ \mbox{MeV}
\vspace*{0.3cm} \\
\tau_{\ osc} \ = \ ( \ 2 \pi \ ) \ / \ \Delta \ m 
\ \sim \ 10^{\ 20} \ \mbox{sec} 
\ = \ 3.3 \ . \ 10^{\ 12} \ \mbox{y}
\end{array}
\end{equation}

\hspace*{-0.2cm} ``Erstens kommt es anders, zweitens als man
denkt.''\footnote{\hspace*{0.1cm} Not only this is clearly unobservable, but
eq. (14) ignores CP violation (no h.o.) , and details of neutrino mass and
mixing, which induces $\Delta \ m$ in eqs. (14-15).}  (``First it happens
differently, second as one thinks.'')

\section{From mesonium to neutrino's $\left \lbrack \ 14 \ \right \rbrack$
\footnote{\hspace*{0.1cm} 
$\left \lbrack \ 14 \ \right \rbrack$
Bruno Pontecorvo, ``Inverse $\beta$ processes
and nonconservation of lepton charge'', JETP (USSR) 34 (1957) 247,
english translation Soviet Physics, JETP 7 (1958) 172. 
}}

\noindent
What is to be remembered from ref. $\left \lbrack \ 14 \ \right \rbrack$
is the  i d e a  of neutrino oscillations, expressed in the 
{\it corrected} sentence :
\vspace*{0.1cm}

\noindent
``The effects due to {\it neutrino flavor transformations} 
 may not be 
observable in the laboratory, owing to the large R, but they will
take place on an astronomical scale.''
\vspace*{0.1cm}

\noindent
The $( \ V \ - \ A \ )  \times  ( \ V \ - \ A \ )$ form
of the Fermi interaction $\left \lbrack \ 15 \ \right \rbrack$
\footnote{\hspace*{0.1cm} $\left \lbrack \ 15 \ \right \rbrack$ Richard
Feynman and Murray Gell-Mann, ``Theory of Fermi interaction'', Phys.\ Rev.\
109 1. January (1958) 193, (no h.o.).}, which subsequently clarified the
structure of neutrino emission and absorption, was {\it documented} almost
contemporaneously.

\section{\boldmath $\Delta \ m \ \tau$ \unboldmath  
from rest system to beam system
$\left \lbrack \ 16 \ \right \rbrack$
\footnote{\hspace*{0.1cm} $\left \lbrack \ 16 \ \right \rbrack$ In notes to
Jack Steinberger, lectures on ``Elementary particle physics'', ETHZ, Zurich
WS 1966/67, not documented (again zigzag in time).}}

\noindent
There is time dilatation from rest system to beam system,
and also we express time in the beam system by 
distance ( $c \ = \ 1$ )

\vspace*{-0.1cm}
\begin{equation}
\label{nu's:16}
\begin{array}{l}
\tau \ \rightarrow 
\ \begin{array}{c}
d
\vspace*{0.2cm} \\ \hline \vspace*{-0.4cm} \\
\gamma \ \beta
\end{array}
\hspace*{0.2cm} ; \hspace*{0.2cm}
\gamma^{\ -1} \ = \ \sqrt{\ 1 \ - \ v^{\ 2} \ }
\hspace*{0.2cm} ; \hspace*{0.2cm}
\beta \ = \ v \ 
\end{array}
\end{equation}
\footnote{\hspace*{0.1cm} 
key question $\rightarrow$
which $v$ ? \hspace*{6.0cm} $\rightarrow$}

\noindent
Then we replace $\Delta \ m$ 
, for 12 beam oscillations

\vspace*{-0.0cm}
\begin{equation}
\label{nu's:17}
\begin{array}{l}
\Delta \ m \ = 
\ \begin{array}{c}
\Delta \ m^{\ 2}
\vspace*{0.2cm} \\ \hline \vspace*{-0.4cm} \\
2 \ \left \langle \ m \ \right \rangle
\end{array}
\hspace*{0.2cm} ; \hspace*{0.2cm}
\begin{array}{ccc}
\Delta \ m^{\ 2} & = & m_{\ 1}^{\ 2} \ - \  m_{\ 2}^{\ 2}
\vspace*{0.2cm} \\ 
\left \langle \ m \ \right \rangle & = & \frac{1}{2}
\ ( \ m_{\ 1} \ + \ m_{\ 2} \ )
\end{array}
\end{array}
\end{equation}

\noindent
Thus we obtain , for {\it any} 12 oscillation phenomenon

\vspace*{-0.0cm}
\begin{equation}
\label{nu's:18}
\begin{array}{l}
\Delta \ m \ \tau \ = 
\ \begin{array}{c}
\Delta \ m^{\ 2}
\vspace*{0.2cm} \\ \hline \vspace*{-0.4cm} \\
2 \ \left \langle \ m \ \right \rangle \ \beta \ \gamma
\end{array}
\hspace*{0.2cm} d
\hspace*{0.3cm} ; \hspace*{0.2cm}
 \left \langle \ m \ \right \rangle
\ \ \beta \ \gamma \ =
\ \left \langle \ p \ \right \rangle
\end{array}
\end{equation}

\noindent
It is apparently clear that 
$ \left \langle \ m \ \right \rangle
\ \  \beta \ \gamma \ =
\ \left \langle \ p \ \right \rangle$
represents the average beam momentum, yet this is not really so.
Lets postpone the questions 

\noindent
{\it ( which
$\left \langle \ p \ \right \rangle$ ? - which $d$ ? )}
 . 

\noindent
From eq. (18) it follows

\vspace*{-0.0cm}
\begin{equation}
\label{nu's:19}
\begin{array}{l}
\Delta \ m \ \tau \ = 
\ \begin{array}{c}
\Delta \ m^{\ 2}
\vspace*{0.2cm} \\ \hline \vspace*{-0.4cm} \\
2 \ \left \langle \ p \ \right \rangle 
\end{array}
\hspace*{0.2cm} d
\hspace*{0.2cm} \rightarrow \hspace*{0.2cm}
\cos \ \left (
\ \begin{array}{c}
\Delta \ m^{\ 2}
\vspace*{0.2cm} \\ \hline \vspace*{-0.4cm} \\
 2 \ \left \langle \ p \ \right \rangle 
\end{array}
\hspace*{0.2cm} d
\ \right )
\end{array}
\end{equation} 
\vspace*{0.2cm}

\noindent
The oscillation amplitude in vacuo
(eq. 19) is well known, yet it contains 'subtleties' .

\begin{center}
\vspace*{0.2cm}

$\Delta \ m^{\ 2} \ d \ / 
\ \left ( \ 2 
\ \left \langle \ p \ \right \rangle \ \right )$ :
what means what ?

\end{center}
\vspace*{0.2cm}

\noindent
The {\it semiclassical} intuition from beam dynamics and optical interference
is obvious.  A well collimated and within 
$\Delta \ | \ \vec{p} \ \ | \ / 
\ \left | \ \left \langle \ \vec{p} \ \right \rangle \ \right |$ 
\hspace*{-0.1cm} \footnote{\hspace*{0.1cm} -- or any similar 
definition of beam momentum spread --} 
\hspace*{0.1cm} 'monochromatic' 
beam is considered as a classical line, lets say along the positive z-axis, 
defining the mean direction from a definite production point 
( $\vec{x} \ = \ 0$ )
towards a detector, at distance d . 

\noindent
But the associated operators
for a single beam quantum

\vspace*{-0.0cm}
\begin{equation}
\label{nu's:20}
\begin{array}{l}  
\widehat{p}_{\ z} \ , \ \widehat{z} 
\hspace*{0.3cm} \rightarrow \hspace*{0.3cm}
\Delta \ \widehat{p}_{\ z} \ \Delta \ \widehat{z} \ \geq \ \frac{1}{2} 
\end{array}
\end{equation}

\noindent
are subject to the uncertainty principle ( using units $\hbar \ = \ 1$ ) .
The same is true for energy and time. 
\vspace*{0.2cm}

\noindent
Yet we are dealing 
in oscillations --
 with single quantum interference
 --
and thus the spread from one beam quantum to the next is only
yielding a 'good guess' of the actual expectation values, e.g. appearing in
eq. (20) .
\vspace*{0.2cm}
 
\noindent
The quantity $\left \langle \ p \ \right \rangle  $ in the expression
for the phase

\begin{equation}
\label{nu's:21}
\begin{array}{l}
 \Delta \ m^{\ 2} \ d \ / 
\ \left ( \ 2 
\ \left \langle \ p \ \right \rangle \ \right )
\end{array}
\end{equation}

\noindent
{\it  essentially presupposes} 
the single quantum production wave function, e.g. in 3 momentum space
in a given fixed frame, propagating from a production time $t_{\ P}$
to a specific detection space-time point $x_{\ D}$
and characterized accounting for {\it all quantum mechanical 
uncertainties} by the distance d . 
In this framework $\left \langle \ p \ \right \rangle$ stands for
the so evaluated single quantum expectation value 
\footnote{\hspace*{0.1cm}
This was the content of my notes in ref. $\left \lbrack \ 16 \ \right \rbrack$
(1966) . \hspace*{1.5cm} h.o. $\rightarrow$ } .
\vspace*{0.2cm}

\noindent
This was {\it implicit} in refs. (e.g.) 
$\left \lbrack \ 5 \ \right \rbrack \ - \ \left \lbrack \ 7 \ \right \rbrack$,
and became obvious in discussing matter effects, specifically
for neutrino oscillations (e.g. in the sun) .
\vspace*{0.2cm}

\section{Coherence and decoherence in neutrino oscillations (h.o.)}

\noindent
The ensuing is an {\it incomplete} attempt of a historical overview,
going zigzag in time, starting with
ref. $\left \lbrack \ 17 \ \right \rbrack$ 
\footnote{\hspace*{0.1cm} 
$\left \lbrack \ 17 \ \right \rbrack$
Carlo Giunti, ``Theory of neutrino oscillations'', hep-ph/0409230,
 in itself a h.o. 
}.  

\noindent
Just mention is due to two papers : 
Shalom Eliezer and Arthur Swift $\left \lbrack \ 18 \ \right \rbrack$ 
\footnote{\hspace*{0.1cm}
$\left \lbrack \ 18 \ \right \rbrack$ Shalom Eliezer and Arthur Swift,
``Experimental consequences of $\nu_{\ e} \ - \ \nu_{\ \mu}$ mixing
in neutrino beams'', Nucl. Phys. B105 (1976) 45,
 submitted 28. July 1975.
}
and Samoil Bilenky and Bruno Pontecorvo
$\left \lbrack \ 19 \ \right \rbrack$ 
\footnote{\hspace*{0.1cm}
$\left \lbrack \ 19 \ \right \rbrack$ Samoil Bilenky and Bruno Pontecorvo, 
``The lepton-quark analogy and muonic charge'', Yad. Fiz. 24 (1976) 603, 
submitted 1. January 1976.} ,
where the phase argument 
$ \Delta \ m^{\ 2} \ d \ / 
\ \left ( \ 2 
\ \left \langle \ p \ \right \rangle \ \right )$ 
appears correctly. 
Also in 1976 a contribution by Shmuel Nussinov 
$\left \lbrack \ 20 \ \right \rbrack$ appeared
\footnote{\hspace*{0.1cm}
$\left \lbrack \ 20 \ \right \rbrack$ 
Shmuel Nussinov, ``Solar neutrinos and neutrino mixing'',
Phys.Lett.B63 (1976) 201, submitted 10. May 1976. } .

\section{Matter effects -- MSW for neutrinos 
[ 21 ] \footnote{\hspace*{0.1cm} $\left \lbrack \ 21 \ \right \rbrack$
Lincoln Wolfenstein, ``Neutrino oscillations in matter'', Phys.\ Rev.\ D17
(1978) 2369.}, [ 22 ] \footnote{\hspace*{0.1cm} $\left \lbrack \ 22 \ \right
\rbrack$ Stanislav Mikheyev and Alexei Smirnov, Sov.\ J.\ Nucl.\ Phys.\ 42
(1985) 913.}}

\noindent
The general remark hereto is 
\vspace*{0.2cm}

\noindent
``Every conceivable coherent or incoherent phenomenon involving 
photons, is bound to happen (and more) with neutrinos.'' 

{\it \hspace*{1.0cm} $\rightarrow$ 
refraction, double refraction, \v{C}erenkov radiation,  $\cdots$
$\left \lbrack \ 23 \ \right \rbrack$
\footnote{\hspace*{0.1cm}
$\left \lbrack \ 23 \ \right \rbrack$
see e.g. Arnold Sommerfeld, ``Optik'', ``Elektrodynamik'', 
``Atombau und Spektrallinien'', 
Akademische Verlagsgesellschaft, Geest und Ko., Leipzig 1959.} 
 .}
\vspace*{0.2cm}

\noindent
The forward scattering amplitude and refractive index relation
is \\ 
(a semiclassical one) \hspace*{2.0cm} $\rightarrow$

\begin{center}
\vspace*{-0.0cm}

plane wave distortion in the z-direction

\end{center}

\vspace*{-0.0cm}
\begin{equation}
\label{nu's:22}
\begin{array}{l}
f_{\ 0} \ \equiv \ f_{\ forward}^{\ lab} \ = 
\ ( \ 8 \pi \ m_{\ target} \ )^{\ -1} \ T_{\ forward}
\vspace*{0.3cm} \\
Im \ f_{\ 0} \ = 
\ \left ( 
\ k_{\ lab} \ / \ ( 4 \pi \ )
\ \right ) \ \sigma_{\ tot}
\hspace*{0.2cm} ; \hspace*{0.2cm} 
k_{\ lab} \ \rightarrow \ k
\vspace*{0.3cm} \\
\begin{array}{lll}
e^{\ i \ k \ z} \ \rightarrow \ e^{\ i \ n \ k \ z}
& = & e^{\ i \ k \ z}
\ e^{\ i \ ( \ 2 \pi \ / \ k \ ) \ \varrho_{\ N} \ f_{\ 0} \ z}
\vspace*{0.3cm} \\
 &  = &  e^{\ i \ k \ z}
\ e^{\ i \ \left \lbrack \ ( \ 2 \pi \ / \ k^{\ 2} \ ) 
\ \varrho_{\ N} \ f_{\ 0} \ \right \rbrack \ k \ z}
\end{array}
\vspace*{0.2cm} \\
n \ = \ 1 \ + 
\ ( \ 2 \pi \ / \ k^{\ 2} \ ) 
\ \varrho_{\ N} \ f_{\ 0}
\hspace*{0.2cm} ; \hspace*{0.2cm}
\left \langle \ v \ \right \rangle_{\ mat.} \ \sim \ 1 \ / \ ( \ Re \ n \ )
 \stackrel{\hspace*{-0.3cm} <}{> \mbox{\small ?}} \ 1 
\vspace*{0.3cm} \\
\hspace*{1.0cm}
\varrho_{\ N} \ = \ \mbox{mean number density of 
(target-) matter}
\vspace*{0.3cm} \\
\hspace*{1.0cm}
T \ = \ \mbox{invariantly normalized (elastic-) scattering amplitude}
\end{array}
\end{equation}
\footnote{\hspace*{0.1cm} key question $\rightarrow$
which is the fully quantum mechanical 
description ?}
\vspace*{-0.0cm}

\noindent
for neutrinos $\left \lbrack \ 24 \ \right \rbrack$ 
\footnote{\hspace*{0.1cm}
$\left \lbrack \ 24 \ \right \rbrack$
Hans Bethe, ``A possible explanation of the solar neutrino puzzle'',
Phys.Rev.Lett.56 (1986) 1305, {\it indeed} , 
tribute --  to many who 'really did it' -- and to a pioneer of 
solar physics and beyond (no h.o.) . }
at low energy :

\vspace*{-0.0cm}
\begin{equation}
\label{nu's:23}
\begin{array}{l}
{\cal{H}}_{\ \nu} \ \sim \ 2 \sqrt{2} \ G_{\ F}
\ \left (
\ \begin{array}{c} 
\overline{\nu}_{\ \alpha} \ \gamma^{\ \mu}_{\ L} \ \nu_{\ \beta}
\ \ \overline{\ell}_{\ \beta} \ \gamma_{\ \mu \ L} \ \ell_{\ \alpha}
\vspace*{0.3cm} \\
+  
\ \overline{\nu}_{\ \alpha} \ \gamma^{\ \mu}_{\ L} \ \nu_{\ \alpha} 
\ j_{\ \mu \ n} \ ( \ \ell \ , \ q \ )
\ \varrho
\vspace*{0.3cm} \\
 + 
\ \frac{1}{4} \ \overline{\nu}_{\ \alpha} \ \gamma^{\ \mu}_{\ L} 
\ \nu_{\ \alpha}
\ \ \overline{\nu}_{\ \beta} \ \gamma^{\ \mu}_{\ L} \ \nu_{\ \beta}
\ \varrho
\end{array}
\ \right )
\vspace*{0.4cm} \\
\alpha \ , \ \beta \ = \ I,II,III \ \mbox{for family}
\hspace*{0.1cm} ; \hspace*{0.1cm}
\varrho \ : \ \mbox{e.w. neutral current parameter}
\end{array}
\end{equation}

\noindent
The second and third \footnote{\hspace*{0.1cm}
the latter induces -- tiny -- matter distortions on relic neutrinos ,
 maybe it is worth while to work them out ?}
terms on the r.h.s. of eq. (23) 
-- in matter consisting of hadrons and electrons --
do not distinguish 
between neutrino flavors. So the relative distortion of $\nu_{\ e}$ 
--  by electrons at rest  -- is

\vspace*{-0.0cm}
\begin{equation}
\label{nu's:24}
\begin{array}{l}
\Delta_{\ e} \ {\cal{H}}_{\ \nu} \ \rightarrow
\ \sqrt{2} \ G_{\ F} \ \nu^{\ \dot{\beta} \ *}_{\ e} \ \nu^{\ \dot{\beta}}_{\ e}
\ \left \langle \ e^{\ *} \ e \ \right \rangle_{\ e}  
\vspace*{0.3cm} \\
\hspace*{2.0cm} 
= \ \left ( \ \sqrt{2} \ G_{\ F} \ \varrho_{\ n_{\ e}} \ \right )
\ \nu^{\ \dot{\beta} \ *}_{\ e} \ \nu^{\ \dot{\beta}}_{\ e}
\vspace*{0.3cm} \\
K_{\ e} \ = \ \sqrt{2} \ G_{\ F} \ \varrho_{\ n_{\ e}}
\hspace*{0.1cm} ; \hspace*{0.1cm}
( \ K_{\ e} \ \rightarrow \ - \ K_{\ e} \ \mbox{for}
\ e^{\ -} \ \rightarrow \ e^{\ +} \ )
\end{array}
\end{equation}
\vspace*{0.1cm}

\noindent
The spinor {\it field} equation 
in the above semiclassical approximation 
in chiral basis becomes -- suppressing all indices --
and allowing for an $\vec{x}$ dependent electron density

\vspace*{-0.0cm}
\hspace*{0.3cm}
\begin{equation}
\label{nu's:25}
\begin{array}{l}
\begin{array}{l}
\left ( \ i \ \partial_{\ t} \ - \ \kappa \ \right ) 
\ \nu \ = \ {\cal{M}}^{\ \dagger} 
\ \widetilde{\ \nu} 
\vspace*{0.3cm} \\
\left ( \ i \ \partial_{\ t} \ + \ \kappa \ \right )
\ \widetilde{\nu} \ = \ {\cal{M}} 
\ \nu
\end{array}
\hspace*{0.1cm} ; \hspace*{0.1cm} i \ \partial_{\ t} \ \rightarrow \ E
\vspace*{0.3cm} \\
\kappa \ = \ K_{\ e} \ P_{\ e} \ - \ \frac{1}{i} 
\ \vec{\sigma} \ \vec{\nabla} \ \P
\hspace*{0.1cm} ; \hspace*{0.1cm} 
\nu \ = \ \nu^{\ \dot{\beta}}_{\ f}
\hspace*{0.2cm} ; \hspace*{0.1cm} 
\widetilde{\nu} \ = \ \varepsilon_{\ \alpha \ \beta}
\ \left ( \ \nu^{\ \dot{\beta}}_{\ f} \ \right )^{\ *}
\vspace*{0.3cm} \\
f \ = \ 1,\cdots,6
\hspace*{0.2cm} ; \hspace*{0.1cm}
P_{\ e} \ = \ \delta_{\ e f} \ \delta_{\ e f'}
\hspace*{0.2cm} ; \hspace*{0.1cm}
\P \ = \ \delta_{\ f f'}
\hspace*{0.2cm} ; \hspace*{0.1cm}
{\cal{M}} \ = \ {\cal{M}}_{\ f f'}
\vspace*{0.3cm} \\
\left ( \ \nu \ , \ \widetilde{\nu} \ \right ) 
\ ( \ t \ , \ \vec{x} \ )
\ : \ \mbox{fields}
\hspace*{0.2cm} ; \hspace*{0.1cm}
^{\ *} \ : \ \mbox{hermitian conjugation}
\vspace*{0.3cm} \\
K_{\ e} \ = \ K_{\ e} \ ( \ \vec{x} \ ) \ =
\ \sqrt{2} \ G_{\ F} \ \ \varrho_{\ n_{\ e}} \ ( \ \vec{x} \ ) 
\end{array}
\end{equation}

\noindent
Here we substitute {\it fields} by wave functions \footnote{\hspace*{0.1cm}
The quantities $\widetilde{\nu} \ ( \ E \ , \ \vec{x} \ )$ are not related
to complex conjugate entries for $\nu \ ( \ E \ , \ \vec{x} \ )$ .}

\vspace*{-0.0cm}
\begin{equation}
\label{nu's:26}
\begin{array}{l}
\left ( \ \nu \ , \ \widetilde{\nu} \ \right ) 
\ \rightarrow \left . \ e^{\ - i E t}
\ \left ( \ \nu \ , \ \widetilde{\nu} \ \right ) \ ( \ E \ , \ \vec{x} \ )
\hspace*{0.3cm} \right |
\vspace*{0.3cm} \\
\rightarrow \ 
\left ( \ E \ - \ \kappa \ \right ) 
\ \nu \ = \ {\cal{M}}^{\ \dagger} \ \widetilde{\nu} 
\hspace*{0.2cm} ; \hspace*{0.1cm}
\left ( \ E \ + \ \kappa \ \right )
\ \widetilde{\nu} \ = \ {\cal{M}} \ \nu
\end{array}
\end{equation}

\noindent
From eq. (26) we obtain the 'squared' form \footnote{\hspace*{0.1cm} Most of
the quantities in the last two lines of eq. (27) give (e.g. in the sun)
negligible effects for the light flavors
\hspace*{0.1cm} $\rightarrow \ 0$.}

\vspace*{-0.0cm}
\begin{equation}
\label{nu's:27}
\begin{array}{l}
\begin{array}{lll}
\left ( \ E^{\ 2} \ - \ \kappa^{\ 2} \ -  
\ {\cal{M}}^{\ \dagger} \ {\cal{M}}
\ \right ) 
\ \nu & = &  
\left \lbrack \ \kappa \ , \ {\cal{M}}^{\ \dagger} \ \right \rbrack
\ \widetilde{\ \nu} 
\vspace*{0.3cm} \\
\left ( \ E^{\ 2} \ - \ \kappa^{\ 2} \ -
\ {\cal{M}} \ {\cal{M}}^{\ \dagger}
 \ \right )
\ \widetilde{\nu} &  = & 
\ \left \lbrack \ {\cal{M}} \ , \ \kappa  \hspace*{0.2cm} 
\right \rbrack
\ \nu
\end{array}
\vspace*{0.2cm} \\
\left \lbrack \ {\cal{M}} \ , \ \kappa \ \right \rbrack
\ = \ K_{\ e} 
\ \left \lbrack \ {\cal{M}} \ , \ P_{\ e} \ \right \rbrack
\hspace*{0.2cm} ; \hspace*{0.2cm}
\left .
\left \lbrack \ \kappa \ , \ {\cal{M}}^{\ \dagger} \ \right \rbrack
\ = \ \left \lbrack \ {\cal{M}} \ , \ \kappa \ \right \rbrack^{\ \dagger}
\hspace*{0.3cm} \right |
\vspace*{0.2cm} \\
\kappa^{\ 2} \ = \ - \ \Delta \ - \ 2 \ P_{\ e} \ K_{\ e} \ \frac{1}{i}
\ \vec{\sigma} \ \vec{\nabla}
\ - \ P_{\ e} \ \vec{\sigma} 
\ ( \ \frac{1}{i} \ \vec{\nabla} \ K_{\ e} \ )
\ + \ P_{\ e} \ K_{\ e}^{\ 2}
\end{array}
\end{equation}

\noindent
In eqs. (26-27) the {\it mainly} neutrinos have
negative helicity, which can be specified precisely if
the small quantities are ignored,
whereas the {\it mainly} antineutrinos
carry positive helicity, in the ultrarelativistic limit
$p \ = \ \left | \ \vec{p} \ \right | \ \gg \ | \ m \ |$ . 

\vspace*{-0.0cm}
\begin{equation}
\label{nu's:28}
\begin{array}{l}
\kappa^{\ 2} \ \rightarrow \ p^{\ 2} \ \pm \ 2 \ p \ K_{\ e} \ P_{\ e}
\ : \ \left \lbrace
\ \begin{array}{l}
+ \hspace*{0.1cm} \mbox{for neutrinos}
\vspace*{0.2cm} \\
- \hspace*{0.1cm} \mbox{for antineutrinos}
\ \end{array} \right .
\end{array}
\end{equation}

\noindent
The mass diagonalization yields correspondingly for the mixing in vacuo
{\it approximatively} 

\vspace*{-0.0cm}
\begin{equation}
\label{nu's:29}
\begin{array}{l}
{\cal{M}}^{\ \dagger} \ {\cal{M}} \ \rightarrow
\ \overline{u} \ m_{\ diag}^{\ 2} \ \overline{u}^{\ -1}
\hspace*{0.2cm} ; \hspace*{0.2cm}
{\cal{M}} \ {\cal{M}}^{\ \dagger} \rightarrow
\ u \ m_{\ diag}^{\ 2} \ u^{\ -1} 
\end{array}
\end{equation}

\noindent
In eq. (29) all quantities are projected 
onto the three light flavors \footnote{\hspace*{0.1cm}
So for real (i.e. orthogonal) u , neutrinos distorted by electrons react
identically to antineutrinos relative to positrons .}.

\newpage
\begin{center}
 \begin{figure}[htb]
\epsfig{file=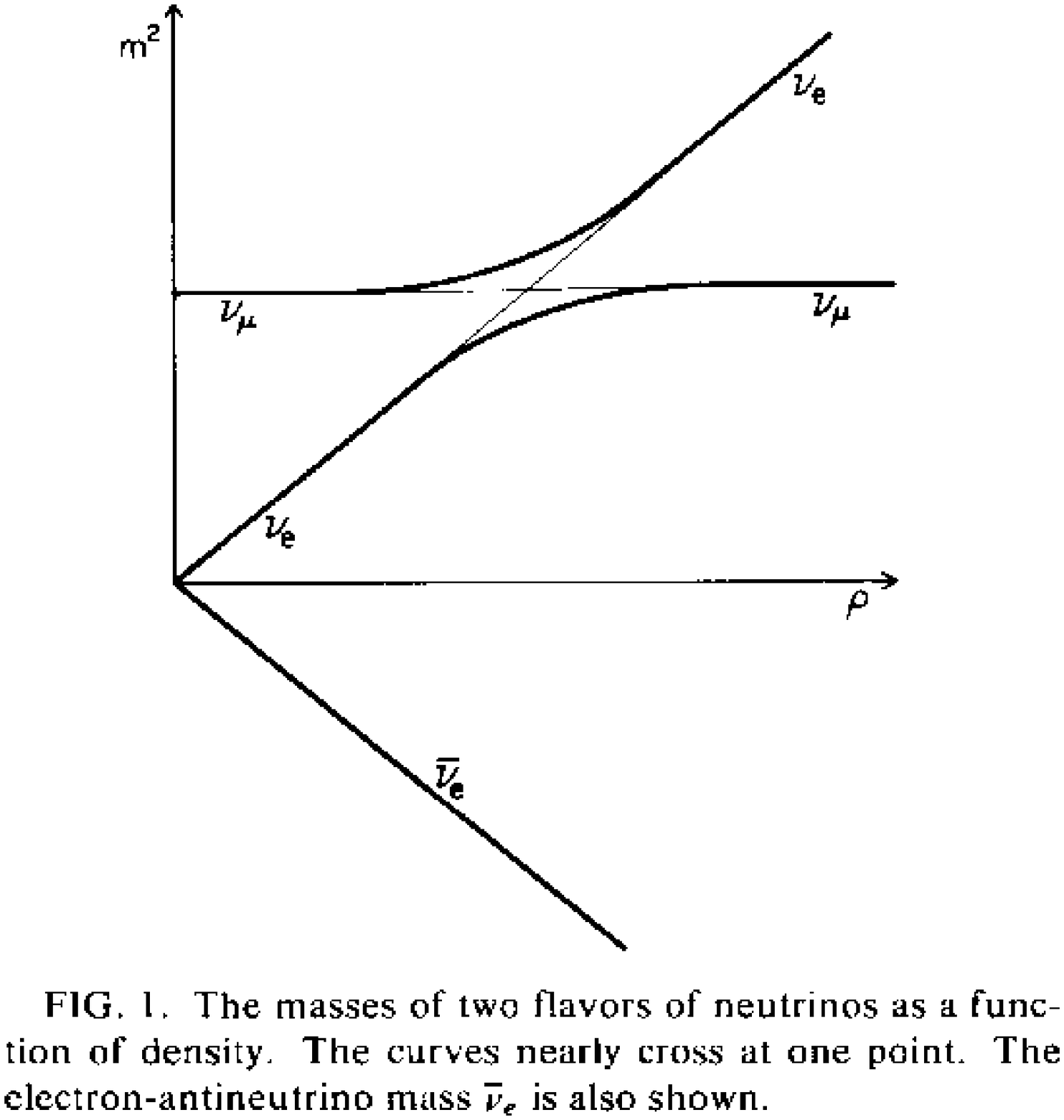,width=12cm}
\vskip -0.2cm
 \end{figure}
\vspace*{-2cm} \footnote{\hspace*{0.1cm}
From Hans Bethe, ref.\ $\left \lbrack \ 24 \ \right \rbrack$, with apologies
to Mikheyev and Smirnov and many.}
\end{center}

\section{(Further) references to the $\odot$ LMA solution
\footnote{\hspace*{0.1cm} See also many refs. 
cited therein, (no h.o.).}}

\noindent
$\left \lbrack \ 25 \ \right \rbrack$ Alexei Smirnov, 
``The MSW effect and matter effects 
in neutrino oscillations'', hep-ph/0412391. 

\vspace*{-0.3cm}
\begin{equation}
\label{nu's:30}
\begin{array}{l}
\Delta \ m^{\ 2}_{\ \odot} \ = \ 7.9 \ . \ 10^{\ -5} \ \mbox{ev}^{\ 2}
\hspace*{0.2cm} ; \hspace*{0.2cm}
\tan^{\ 2} \ \vartheta_{\ \odot} \ = \ 0.40 
\hspace*{0.2cm} ; \hspace*{0.2cm} 
\vartheta \ \sim \ 32.3^{\ \circ}
\end{array}
\end{equation}

\noindent
$\left \lbrack \ 26 \ \right \rbrack$ Serguey Petcov, 
``Towards complete neutrino mixing matrix and CP-violation'', hep-ph/0412410.

\vspace*{-0.3cm}
\begin{equation}
\label{nu's:31}
\begin{array}{l}
\Delta \ m^{\ 2}_{\ \odot} \ = \ \left ( \ 7.9^{\ +0.5}_{\ -0.6} 
\ \right ) \ . \ 10^{\ -5} \ \mbox{ev}^{\ 2}
\hspace*{0.2cm} ; \hspace*{0.2cm}
\tan^{\ 2} \ \vartheta_{\ \odot} \ = \ 0.40^{\ +0.09}_{\ -0.07} 
\hspace*{0.2cm} \mbox{Fig.} \hspace*{0.2cm}  \rightarrow
\end{array}
\end{equation} 
 
\noindent
$\left \lbrack \ 27 \ \right \rbrack$ John Bahcall and Carlos Pena-Garay, 
``Global analyses as a road map to solar neutrino fluxes and 
\hspace*{0.6cm} oscillation parameters'',
JHEP 0311 (2003) 004, hep-ph/0305159. 

\noindent
$\left \lbrack \ 28 \ \right \rbrack$
Gian Luigi Fogli, Eligio Lisi, Antonio Marrone, Daniele Montanino, 
Antonio Palazzo and A.M. \hspace*{0.6cm} Rotunno, 
``Neutrino oscillations: a global analysis'', hep-ph/0310012.

\noindent
$\left \lbrack \ 29 \ \right \rbrack$
P. Aliani, Vito Antonelli, Marco Picariello and Emilio Torrente- Lujan, 
``The neutrino mass matrix \hspace*{0.6cm} after Kamland and 
SNO salt enhanced results'', hep-ph/0309156.

\noindent
$\left \lbrack \ 30 \ \right \rbrack$
Samoil Bilenky, Silvia Pascoli and  Serguey Petcov, 
``Majorana neutrinos, neutrino mass spectrum, 
\hspace*{0.6cm} CP violation and 
neutrinoless double beta decay. 1. The three neutrino mixing case'',
Phys.Rev.D64 \hspace*{0.6cm} (2001) 053010, hep-ph/0102265.

\vspace*{-0.2cm}
\begin{center}
 \begin{figure}[htb]

\epsfig{file=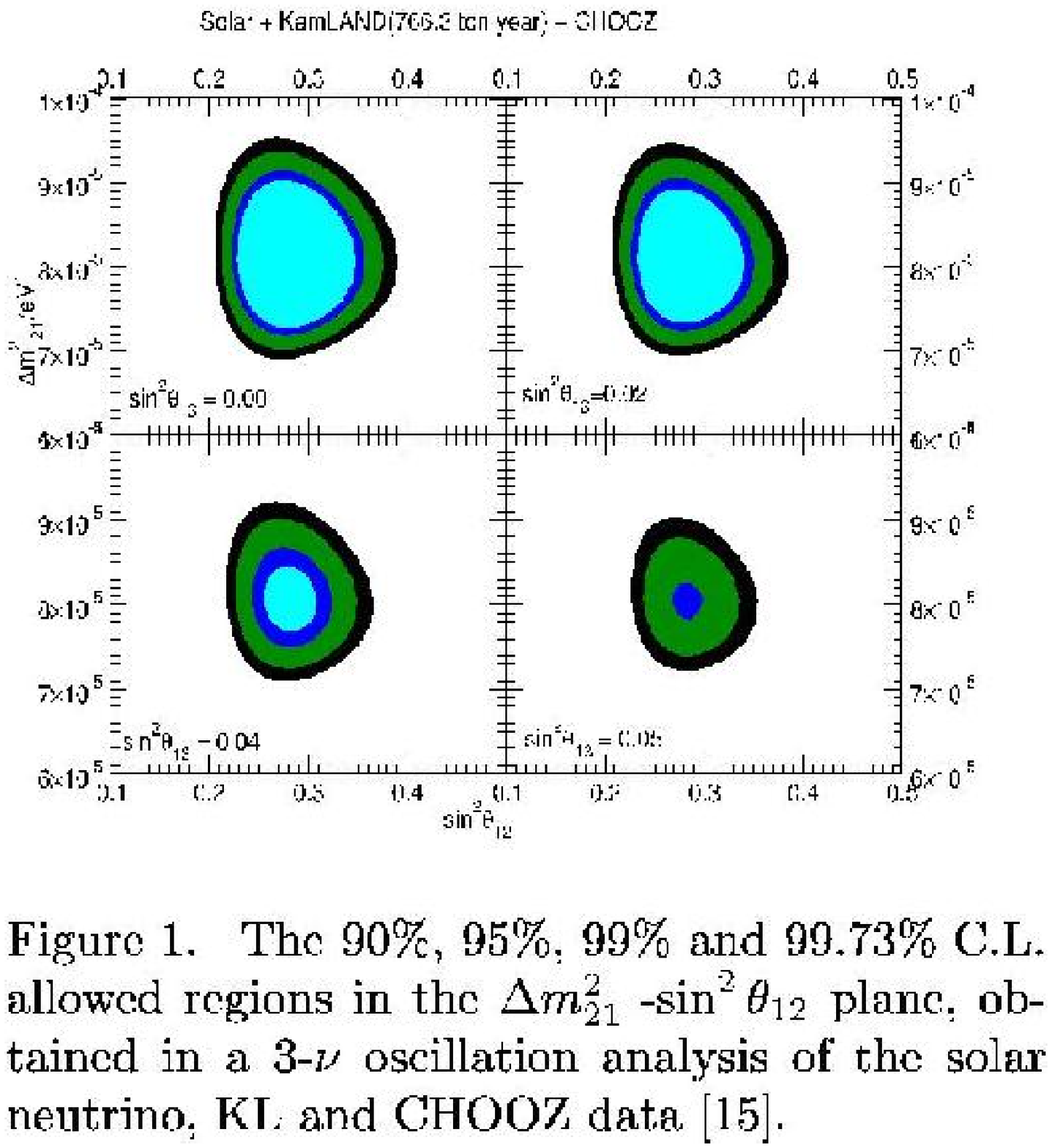,width=11.cm}
\vskip -0.2cm
 \end{figure}
\end{center}

\section{Experimental references to the $\odot$ LMA solution
\footnote{\hspace*{0.1cm} See also many refs. 
cited therein, (no h.o.).}}

\noindent
$\left \lbrack \ 31 \ \right \rbrack$
Bruce Cleveland, Timothy Daily, Raymond Davis, James Distel, Kenneth Lande, 
Choon-kyu Lee, Paul Wildenhain and Jack Ullman,
``Measurement of the solar electron neutrino flux 
with the Home-stake chlorine 
detector'', 
Astrophys. J. 496 (1998) 505.

\hspace*{2.0cm} reaction : 
$\nu_{\ e \ \odot} \ + \ ^{37}Cl \ \rightarrow \ ^{\ 37}Ar \ + \ e^{\ -}$
\vspace*{0.2cm}

\noindent
$\left \lbrack \ 32 \ \right \rbrack$
V. Gavrin, ``Results from the Russian American gallium experiment'',
for the SAGE collaboration, 
J.N. Abdurashitov et al.,
J. Exp. Theor. Phys. 95 (2002) 181, astro-ph/0204245.

\noindent
J.N. Abdurashitov, V.N. Gavrin, S.V. Girin, V.V. Gorbachev, 
P.P. Gurkina, T.V. Ibragimova, 
A.V. Kalikhov, N.G. Khairnasov, T.V. Knodel, 
I.N. Mirnov, A.A. Shikhin, E.P. Veretenkin, V.M. Vermul, 
V.E. Yants, G.T. Zatsepin,
Moscow, INR

\noindent
T.J. Bowles, W.A. Teasdale,
Los Alamos

\noindent
J.S. Nico,
NIST, Wash., D.C.

\noindent
B.T. Cleveland, S.R. Elliott, J.F. Wilkerson,
Washington U., Seattle. 

\hspace*{2.0cm} reaction :
$\nu_{\ e \ \odot} \ + \ ^{71}Ga \ \rightarrow \ ^{\ 71}Ge \ + \ e^{\ -}$
\hspace*{2.0cm} $\rightarrow$

\noindent
$\left \lbrack \ 33 \ \right \rbrack$ W. Hampel et al.,
 GALLEX collaboration, 
``GALLEX solar neutrino observations: 
results for \hspace*{0.6cm} GALLEX IV'', Phys. Lett. B 447 (1999) 127.

\noindent
W. Hampel, J. Handt, G. Heusser, J. Kiko, T. Kirsten, M. Laubenstein, 
E. Pernicka, W. Rau, M. Wojcik, 
Y. Zakharov,
Heidelberg, Max Planck Inst.

\noindent
R. von Ammon, K.H. Ebert, T. Fritsch, D. Heidt, E. Henrich, 
L. Stieglitz, F. Weirich,
Karlsruhe U., EKP

\noindent
Balata, M. Sann, F.X. Hartmann,
Gran Sasso

\noindent
E. Bellotti, C. Cattadori, O. Cremonesi, N. Ferrari, 
E. Fiorini, L. Zanotti,
Milan U. and INFN, Milan

\noindent
M. Altmann, F. von Feilitzsch, R. M\"{o}ssbauer, S. Wanninger,
Munich, Tech. U.

\noindent
G. Berthomieu, E. Schatzman,
Cote d'Azur Observ., Nice

\noindent
I. Carmi, I. Dostrovsky,
Weizmann Inst.

\noindent
C. Bacci, P. Belli, R. Bernabei, S. d'Angelo, L. Paoluzi,
Rome U.,Tor Vergata and INFN, Rome

\noindent
M. Cribier, J. Rich, M. Spiro, C. Tao, D. Vignaud,
DAPNIA, Saclay

\noindent
J. Boger, R.L. Hahn, J.K. Rowley, R.W. Stoenner, J. Weneser,
Brookhaven. 

\hspace*{2.0cm} reaction :
$\nu_{\ e \ \odot} \ + \ ^{71}Ga \ \rightarrow \ ^{\ 71}Ge \ + \ e^{\ -}$
\hspace*{0.5cm} like SAGE .
\hspace*{1.5cm} $\rightarrow$

\noindent
$\left \lbrack \ 34 \ \right \rbrack$ 
Y. Fukuda et al., Kamiokande collaboration
``Solar neutrino data covering solar cycle 22'', 
Phys. \hspace*{0.6cm} Rev. Lett. 77 (1996) 1683.

\noindent
Y. Fukuda, T. Hayakawa, K. Inoue, K. Ishihara, H. Ishino, 
S. Joukou, 
T. Kajita, S. Kasuga, Y. Koshio, 
T. Kumita, K. Matsumoto, M. Nakahata, 
K. Nakamura, K. Okumura, A. Sakai, M. Shiozawa, J. Suzuki, 
Y. Suzuki, T. Tomoeda, Y. Totsuka,
Tokyo U., ICRR

\noindent
K.S. Hirata, K. Kihara, Y. Oyama,
KEK, Tsukuba

\noindent
Masatoshi Koshiba,
Tokai U., Shibuya

\noindent
K. Nishijima, T. Horiuchi,
Tokai U., Hiratsuka

\noindent
K. Fujita, S. Hatakeyama, M. Koga, T. Maruyama, A. Suzuki,
Tohoku U.

\noindent
M. Mori,
Miyagi U. of Education

\noindent
T. Kajimura, T. Suda, A.T. Suzuki,
Kobe U.

\noindent
T. Ishizuka, K. Miyano, H. Okazawa,
Niigata U.

\noindent
T. Hara, Y. Nagashima, M. Takita, T. Yamaguchi,
Osaka U.

\noindent
Y. Hayato, K. Kaneyuki, T. Suzuki, Y. Takeuchi, T. Tanimori,
Tokyo Inst. Tech.

\noindent
S. Tasaka,
Gifu U.

\noindent
E. Ichihara, S. Miyamoto, K. Nishikawa,
INS, Tokyo. 

\hspace*{2.0cm} reaction :
$\nu_{\ e \ \odot} \ + \ e^{\ -} \ \rightarrow \ \nu_{\ e} \ + \ e^{\ -}$
\vspace*{0.2cm}

\noindent
$\left \lbrack \ 35 \ \right \rbrack$ 
Y.Ashie et al., Super-Kamiokande collaboration,
``Evidence for an oscillatory signature in 
atmospheric neutrino oscillation'',
Phys.Rev.Lett.93 (2004) 101801, hep-ex/0404034 and 

\noindent
Y. Suzuki, ``Super-Kamiokande: present and future'',
Nucl.Phys.Proc.Suppl.137 (2004) 5.  

\noindent
M. Goldhaber, Masatoshi Koshiba, J.G. Learned, S. Matsuno, 
R. Nambu, L.R. Sulak, Y. Suzuki, 
R. Svoboda, 
Y. Totsuka,
R.J. Wilkes $\cdots$ \footnote{\hspace*{0.1cm}
\begin{tabular}{l}
``Wer z\"{a}hlt die Seelen, nennt die Namen, \\ 
\hspace*{0.1cm} die gastlich hier zusammenkamen.''
\end{tabular}
\begin{tabular}{l} 
``Who counts the souls, relates the names, \\
\hspace*{0.1cm} who met in piece here for the games.''
\end{tabular} \\
\hspace*{2.5cm} Friedrich Schiller 
}
\vspace*{0.2cm}
 
\hspace*{2.0cm} reactions :
$\nu_{\ e \ \odot} \ + \ e^{\ -} \ \rightarrow \ \nu_{\ e} \ + \ e^{\ -}$

\hspace*{2.0cm} $\begin{array}{l} 
\nu_{\ e, \mu} \\
\overline{\nu}_{\ e, \mu} 
\end{array}
\ + \ H_{\ 2} \ O \ \rightarrow \ e^{\ \mp} \ , \ \mu^{\ \mp} \ + X$
\vspace*{0.2cm}

\hspace*{2.0cm} $^{\ 8}B$ solar neutrino flux :
\vspace*{0.2cm}

\hspace*{2.0cm} \begin{tabular}{cl}
$5.05 \ \left ( \ 1^{\ +0.20}_{\ -0.16} \ \right ) \ . \ 10^{\ - 6} \ / 
\ cm^{\ 2} \ / \ s$ & for BP2000 \\
$5.82 \ ( \ 1 \ \pm \  0.23 \ ) \ . \ 10^{\ -6} \ / \ cm^{\ 2} \ / \ s$
 & for BP2004
\end{tabular}
\hspace*{0.5cm} $\rightarrow$

\vspace*{-0.0cm}
\begin{center}
 \begin{figure}[htb]

\epsfig{file=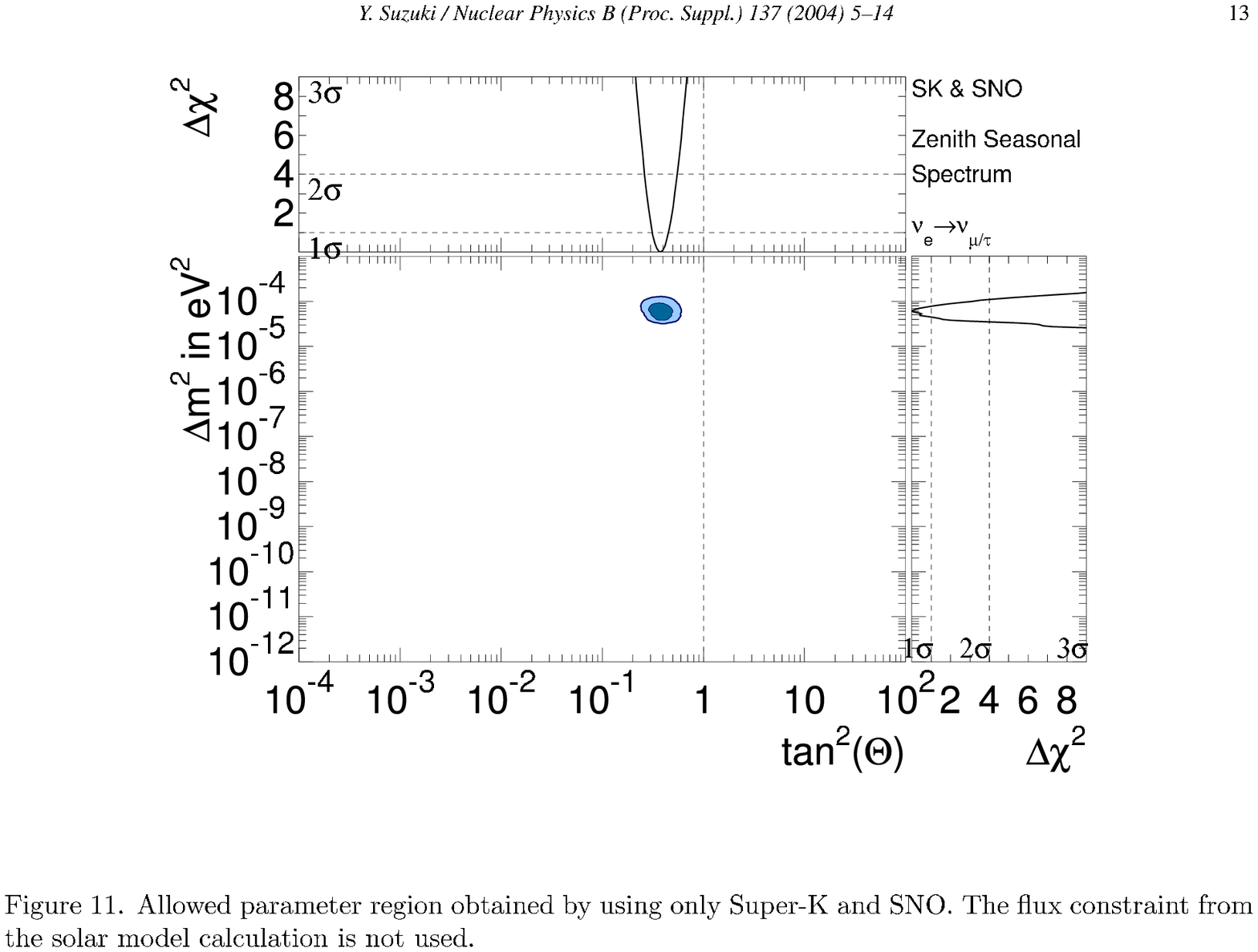,width=12.cm}
\vskip -0.2cm
 \end{figure}
\end{center}

\noindent
$\left \lbrack \ 36 \ \right \rbrack$
D. Sinclair for the SNO collaboration, 
``Recent results from SNO'', \\
\noindent
Nucl.Phys.Proc.Suppl.137 (2004) 150,

\noindent
Art McDonald, A. Hamer, J.J. Simpson, D. Sinclair,
David Wark $\cdots \ ^{\ \leftarrow \ 37}$
\vspace*{0.2cm}

\vspace*{-0.0cm}
\begin{equation}
\label{nu's:32}
\begin{array}{l}
\Delta \ m^{\ 2}_{\ \odot} \ = \ \left ( \ 7.1^{\ +1.0}_{\ -0.6} \ \right ) 
\ . \ 10^{\ -5} 
\ \mbox{ev}^{\ 2}
\hspace*{0.2cm} ; \hspace*{0.2cm}
\vartheta_{\ \odot} \ = \ \left ( \ 32.5^{\ +2.4}_{\ -2.3} 
\ \right )^{\ \circ} 
\end{array}
\end{equation}

\hspace*{2.0cm} reactions :
$\begin{array}[t]{llll}
\nu_{\ e \ \odot} \ + \ d & \rightarrow & p \ + \ p \ + \ e^{\ -} 
 & \mbox{(CC)} \\
\nu_{\ x \ \odot} \ + \ d & \rightarrow & p \ + \ n \ + \ \nu_{\ x} 
 & \mbox{(NC)} \\
\nu_{\ x \ \odot} \ + \ e^{\ -} & \rightarrow & e^{\ -} \ + \ \nu_{\ x}
 & \mbox{(ES)}
\end{array}
$  
\vspace*{0.2cm}

\begin{center}
\vspace*{-0.0cm}

\noindent
Reference(s) to 
$\overline{\nu}_{\ e} \ \leftrightarrow \ \overline{\nu}_{\ x}$
oscillations 

\end{center}
\vspace*{0.2cm}

\noindent
$\left \lbrack \ 37 \ \right \rbrack$
A. Suzuki for the Kamland collaboration,
``Results from Kamland'', \\
Nucl.Phys.Proc.Suppl.137 (2004) 21,

\noindent
T. Araki, K. Eguchi, P.W. Gorham, J.G. Learned, S. Matsuno, 
H. Murayama, Sandip Pakvasa, 
A. Suzuki, R. Svoboda, P. Vogel $\cdots \ ^{\ \leftarrow \ 37}$
\vspace*{0.2cm}

\hspace*{2.0cm} reaction :
\hspace*{0.2cm}
$\overline{\nu}_{\ e \ \odot} \ + \ p \ \rightarrow \ e^{\ -} \ + \ n$
\hspace*{0.2cm}
from $\sim$ 20 reactors

\begin{center}
\vspace*{-0.0cm}

\noindent
leading to the present best estimates 
$\left ( \mbox{ref.} \left \lbrack \ 26 \ \right \rbrack \ \right )$ 
for 3 flavor oscillations :

\end{center}
\vspace*{0.0cm}

\vspace*{-0.0cm}
\begin{equation}
\label{nu's:33}
\begin{array}{l}
\hspace*{0.2cm} \Delta \ m^{\ 2}_{\ \odot} \hspace*{0.25cm} = 
\ \left ( \ 7.9^{\ +0.5}_{\ -0.6} 
\ \right ) \ . \ 10^{\ -5} \ \mbox{ev}^{\ 2}
\hspace*{0.2cm} ; \hspace*{0.2cm}
\tan^{\ 2} \ \vartheta_{\ \odot} \ = \ 0.40^{\ +0.09}_{\ -0.07} 
\vspace*{0.3cm} \\
\left | \ \Delta \ m^{\ 2}_{\ 2 3} \ \right | \ =
\ \left ( \ 2.1_{\ -0.8}^{\ +2.1} \ \right ) \ . \  10^{\ -3} 
\ \mbox{eV}^{\ 2} 
\hspace*{0.2cm} ; \hspace*{0.2cm}
\sin^{\ 2} \ 2 \ \vartheta_{\ 23} \ = \ 1.0_{\ -0.15}  
\vspace*{0.3cm} \\
\hspace*{3.5cm} \sin^{\ 2} \ \vartheta_{\ 13} \ \leq \ 0.05
\hspace*{0.3cm} \mbox{at} \hspace*{0.2cm} 99.73 \ \% \ \mbox{C.L.}
\end{array}
\end{equation}
\vspace*{0.2cm}

\begin{center}
\begin{figure}[h]
\includegraphics[angle=-90,width=11cm]{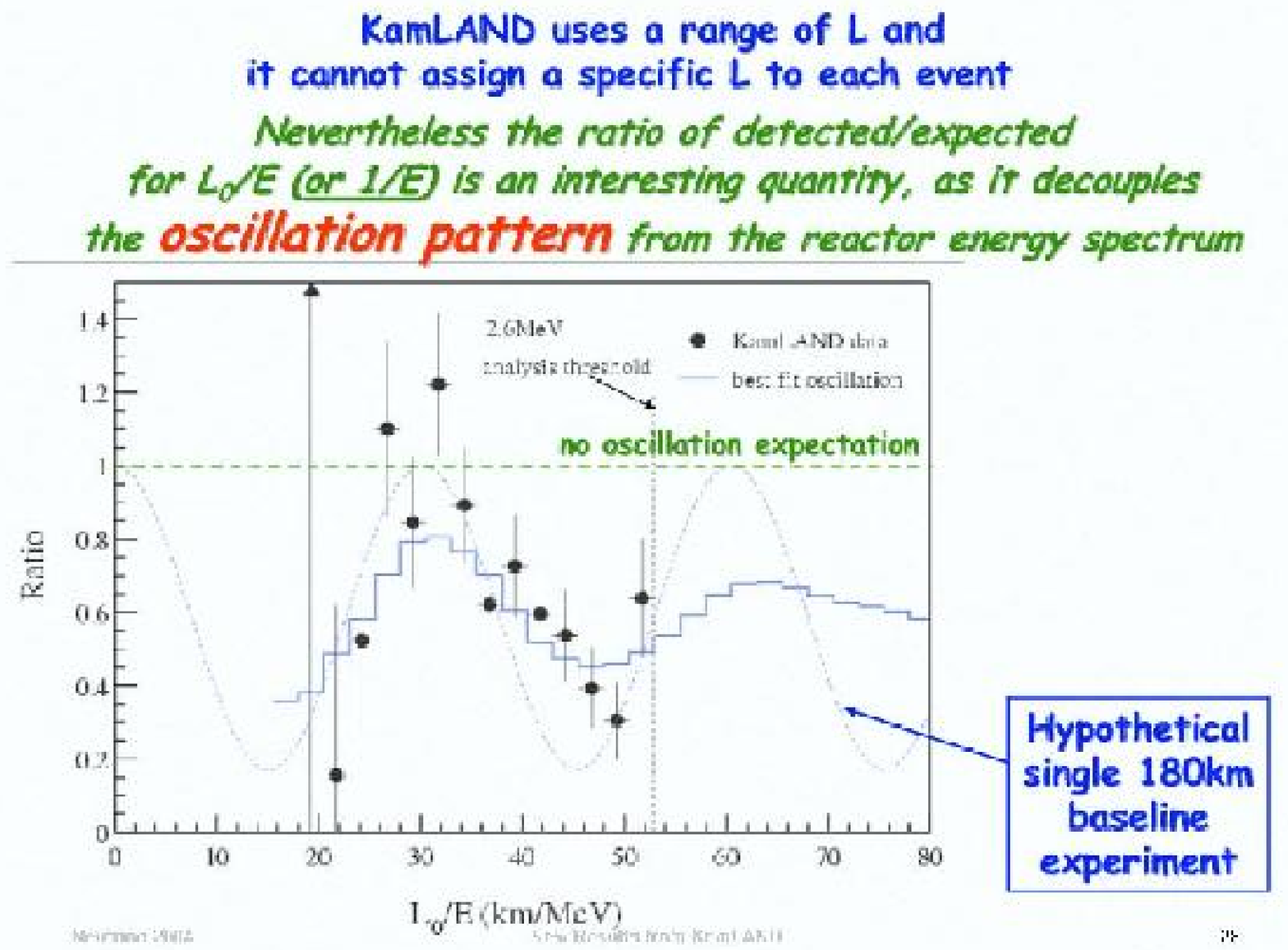}
\end{figure}
\end{center}

\newpage

This shall conclude my -- necessarily partial -- historical overview. What
follows must be cut short.

\section{Mass from mixing $\rightarrow$ the subtle things} 

\noindent
Key questions \hspace*{0.2cm} $\rightarrow$
\hspace*{0.2cm} which is the scale of $M$?
 $O(10^{\ 10})$ GeV  $\rightarrow$
is there any evidence for this scale today ?  hardly !
$\rightarrow$ and what about susy ?

\begin{center}
\begin{figure}[htb]
\epsfig{file=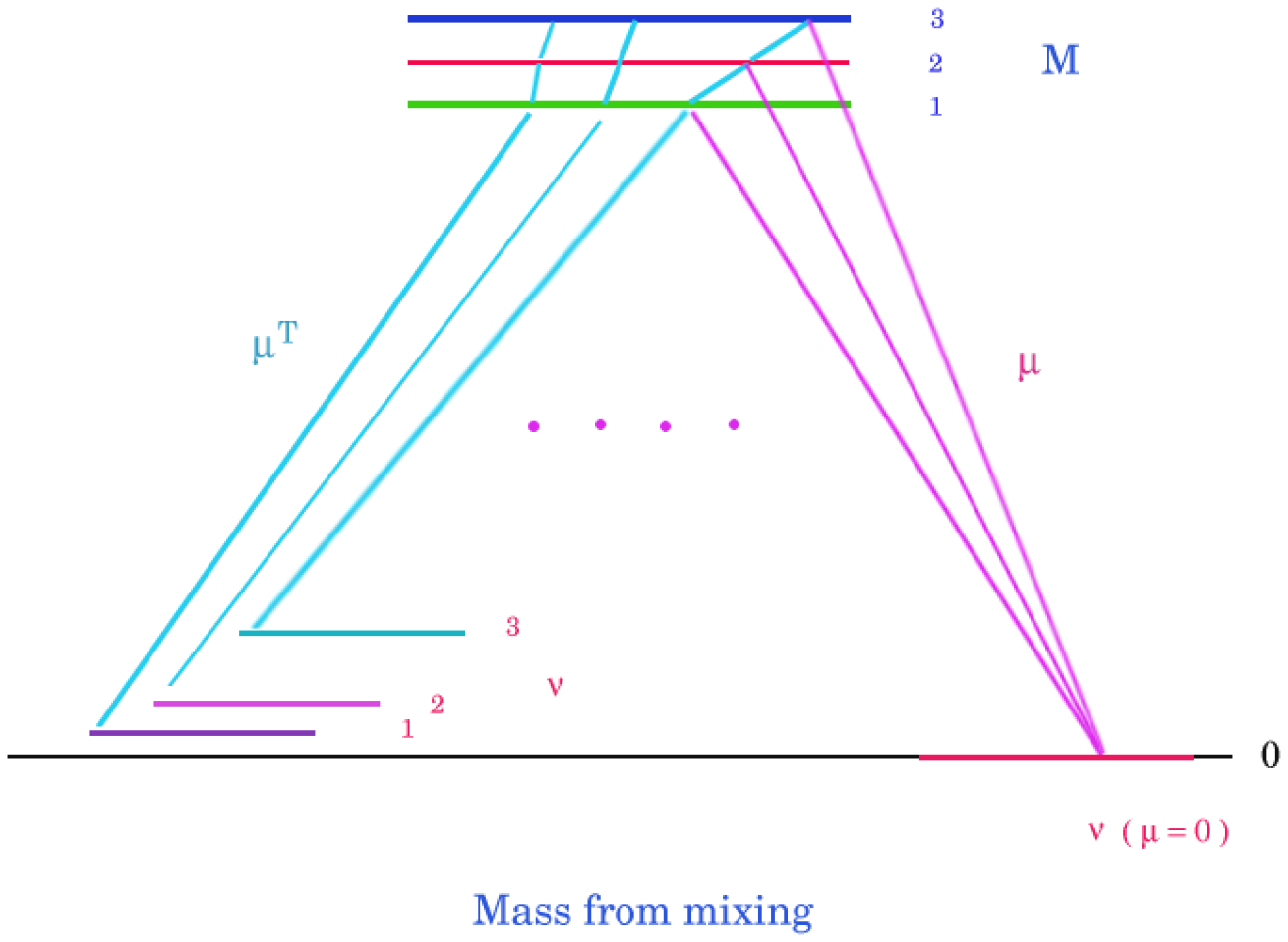,width=10.cm}\vspace*{-0.5cm}
\end{figure}
\end{center}

\noindent
How is the mass matrix of the form in eq. (9) diagonzlized 
{\it exactly} ?

\begin{equation}
\label{nu's:34}
\begin{array}{l}
{\cal{M}} \ =
\ \left (
 \begin{array}{ll}
 0 &  \mu^{\ T}
\vspace*{0.3cm} \\
 \mu & \ M 
\end{array}
  \right )
\ =  
U \ {\cal{M}}_{\ diag} \ U^{\ T}
\hspace*{0.2cm} = \hspace*{0.1cm} 
\ K 
\ {\cal{M}}_{\ bl.diag}
\ K^{\ T}
\vspace*{0.2cm} \\
{\cal{M}}_{\ diag} \ =
{\cal{M}}_{\ diag} \ ( \ m_{\ 1} \ , \ m_{\ 2} \ , \ m_{\ 3} \ ; 
\ M_{\ 1} \ , \ M_{\ 2} \ , \ M_{\ 3} \ )
\vspace*{0.2cm} \\
{\cal{M}}_{\ bl.diag} \ = \ U_{\ 0} \ {\cal{M}}_{\ diag} \ U_{\ 0}^{\ T}
\hspace*{0.2cm} ; \hspace*{0.1cm}
U_{\ 0} \ = 
\ \left (
\begin{array}{cc}
 u_{\ 0} &  0
\vspace*{0.3cm} \\
 0 &  v_{\ 0}
\end{array}
\ \right )
\vspace*{0.3cm} \\
\hspace*{2.5cm}
U \ = \ K \ U_{\ 0} \ = 
\ \left (
\begin{array}{cc}
 u_{\ 11} &   u_{\ 12}
\vspace*{0.3cm} \\
 u_{\ 21} &  u_{\ 22}
\end{array}
\ \right )
\hspace*{1.7cm} \rightarrow
\end{array}
\end{equation}
\vspace*{0.0cm}

\vspace*{-2.6cm}
\begin{equation} 
\label{nu's:35}
\begin{array}{l}
U \ = K U_{\ 0} 
\hspace*{0.1cm} ; \hspace*{0.1cm}
K^{\ -1} \ {\cal{M}} \ K^{\ -1 \ T} \ = \ {\cal{M}}_{\ bl. diag.}
 = 
 \left (
 \begin{array}{cc}
\mbox{\small ${\cal{M}}_{\ 1}$} &  0
\vspace*{0.3cm} \\
 0 & \ {\cal{M}}_{\ 2} 
\end{array}
 \right )
\vspace*{0.3cm} \\ \hline \vspace{-0.3cm} \\
U \ =
\hspace*{0.1cm} 
\left (
 \begin{array}{cc}
\left ( \ 1 \ + \ t \ t^{\ \dagger} \ \right )^{ -1/2}  
\ u_{\ 0}
&  \left ( \ 1 \ + \ t \ t^{\ \dagger} \ \right )^{ -1/2} \ t
\ v_{\ 0}
\vspace*{0.3cm} \\
 - t^{\ \dagger} \ \left ( \ 1 \ + \ t \ t^{\ \dagger} 
\ \right )^{ -1/2} \ u_{\ 0}
 &  \left ( \ 1 \ + \ t^{\ \dagger} \ t \ \right )^{ -1/2} 
\ v_{\ 0} 
\end{array}
 \right )
\vspace*{0.2cm} \\
 \hspace*{6.0cm} \downarrow \ \cdots
\hspace*{3.5cm} \rightarrow
\vspace*{0.2cm} \\
\mbox{${\cal{M}}_{\ 1}$} \ = 
\ - \ t \ {\cal{M}}_{\ 2} \ t^{\ T}
\hspace*{0.1cm} ; \hspace*{0.3cm}
\left \lbrack
\ \begin{array}{llll} 
 \mbox{${\cal{M}}_{\ 1}$} 
&  = &  u_{\ 0} \ m_{\ diag} \ u_{\ 0}^{\ T}
&  \mbox{\bf : \hspace*{0.2cm} light \hspace*{0.1cm} 3}
\vspace*{0.3cm} \\
{\cal{M}}_{\ 2} &  = &  
v_{\ 0} \ M_{\ diag} \ v_{\ 0}^{\ T}
& \mbox{\bf : \hspace*{0.2cm} heavy 3}
\end{array}
\ \right \rbrack
\end{array}
\end{equation}

\noindent
In eq. (35) all matrices are $3 \ \times \ 3$ ,
t describes  light - heavy mixing
{\it generating mass by mixing} .
\footnote{\hspace*{0.1cm} This is {\it documented} 
\hspace*{0.05cm} in $\left \lbrack \ 38 \ \right \rbrack$
Clemens Heusch and Peter Minkowski,
``Lepton flavor violation induced by heavy Majorana neutrinos'' ,
Nucl.Phys.B416 (1994) 3 .  
}
\vspace*{-0.1cm}

\noindent
$u_{\ 0}$ (unitary) accounts for 
light-light mixing  
and $v_{\ 0}$ (unitary) for heavy-heavy (re)mixing .

\noindent
t is 'driven' by $\mu$ ( in ${\cal{M}} \ \leftarrow$ ) and determined 
from the quadratic equation

\vspace*{-0.0cm}
\begin{equation}
\label{nu's:37}
\begin{array}{l}
t \ = \ \mu^{\ T} \ M^{\ -1} - \ t \ \mu \ \overline{t} \ M^{\ -1}
\hspace*{0.1cm} ; \hspace*{0.1cm}
\mbox{ to be solved by iteration $\rightarrow$}
\vspace*{0.3cm} \\
t_{\ n + 1} \ = \ \mu^{\ T} \ M^{\ -1} - 
\ t_{\ n} \ \mu \ \overline{t}_{\ n} \ M^{\ -1}
\hspace*{0.3cm} ; \hspace*{0.3cm}
t_{\ 0} \ = \ 0
\vspace*{0.2cm} \\
t_{\ 1} \ = \ \mu^{\ T} \ M^{\ -1} \ ,
\hspace*{0.2cm} , \hspace*{0.2cm}
t_{\ 2} \ = \ t_{\ 1} \ - 
\ \mu^{\ T} \ M^{\ -1} \ \mu \ \mu^{\ \dagger} \ \overline{M}^{\ -1}
\ M^{\ -1} 
\vspace*{0.1cm} \\
\hspace*{0.5cm}
\cdots
\hspace*{0.2cm} ; \hspace*{0.2cm}
\lim_{\ n \ \rightarrow \ \infty} \ t_{\ n} \ = \ t
\hspace*{0.5cm} 
 \hspace*{2.5cm} \rightarrow
\end{array}
\end{equation}
\vspace*{0.5cm}

\noindent
Finally lets turn to the mixing matrix $u_{\ 11}$ (eqs. 34-35)
\footnote{\hspace*{0.1cm} This is {\it essentially} different from the
mixing of {\it identical} $SU2_{\ L} \ \times \ U1$ representations, i.e.\
charged \hspace*{0.1cm} base fermions.},\footnote{\hspace*{0.1cm}
$\left \lbrack \ 39 \ \right \rbrack$ Ziro Maki, Masami Nakagawa and Shoichi
Sakata, Prog.\ Theor.\ Phys.\ 28 (1962) 870.
\hspace*{0.2cm} $\rightarrow$ the PMNS-matrix: The authors assumed (in 1962)
only light-light mixing.}

\begin{equation}
\label{nu's:38}
\begin{array}{l}
u_{\ 11} \ =
\ \left ( \ 1 \ + \ t \ t^{\ \dagger} \ \right )^{ -1/2}  
\ u_{\ 0}
\ \sim \ u_{\ 0} \ - \ \frac{1}{2} \ t \ t^{\ \dagger} \ u_{\ 0}
\vspace*{0.3cm} \\
t \ t^{\ \dagger} \ = \ O 
\ \left ( \ \overline{m} \  / \ \overline{M}
\ \right )
\ \sim \ 10 \ \mbox{meV}  \ / \ 10^{\ 10} \ \mbox{GeV}
\ = \ 10^{\ -21}
\end{array}
\end{equation}

\noindent
The {\it estimate} in eq. (37) is very uncertain and
assumes among other things $ m_{\ 1} \ \sim \ 1 \ \mbox{meV}$ .

\noindent
Nevertheless it follows on the same grounds as the smallness of light
neutrino masses, that the deviation of $u_{\ 11}$
from  $u_{\ 0}$
is tiny .
\vspace*{0.7cm} 

\hspace*{3.0cm} 
\begin{tabular}{c} ``Much ado about nothing'' , 
William Shakespeare 
\vspace*{0.3cm} \\
\_ \_ \_ \_ \_ \_ \_
\end{tabular} 

~

~

\section{Some perspectives} 

\noindent
1) Neutrino properties are only to a very small extent
open (up to the present) to deductions from oscillation measurements.
\vspace*{0.2cm}

\noindent
2) Notwithstanding this, a significant and admirable experimental effort
paired with theoretical analysis has revealed the main two
oscillation modes. The matter effect due to Mikheev, Smirnov and Wolfenstein
demonstrates another clear form of quantum coherence, over
length scales of the solar radius.
\vspace*{0.2cm}

\noindent
3) Key questions remain to be resolved : are all
(ungauged or gauged) global charge-like quantum numbers violated ?
(B-L) , B , L, individual lepton flavors.
\vspace*{0.3cm}

\noindent
4) SO10 served fine (together with susy or without it)
to guide ideas, but a genuine unification is as remote
as the scales and nature of heavy neutrino flavors, to name only
these.
\vspace*{0.3cm}

\noindent
5) I do hope, that not only this workshop ``Neutrino telescopes in Venice''
will continue to bring new insights, but also that powerful neutrino
telescopes will come into existence, last but not least in the sea.
\vspace*{2.0cm}

\hspace*{1.84cm}  And the quest for unification remains wide open

\end{document}